\documentclass[twocolumn,,superscriptaddress,letter]{revtex4}
\usepackage{amssymb}
\usepackage{color}
\usepackage{graphicx}
\usepackage{dcolumn}
\usepackage{bm}
\usepackage[header,title,page,titletoc]{appendix}
\usepackage{amsmath}
\usepackage{amsfonts}

\providecommand{\U}[1]{\protect \rule{.1in}{.1in}}

\begin{document}

\title{Electronic transport of bilayer graphene with asymmetry line defects}
\author{Xiao-Ming Zhao}
\affiliation{Department of Physics, Beijing Normal University, Beijing 100875, China}
\author{Ya-Jie Wu}
\affiliation{School of Science, Xi'an Technological University, Xi'an 710021, PR China}
\author{Chan Chen}
\affiliation{Department of Physics, Beijing Normal University, Beijing 100875, China}
\author{Ying Liang}
\affiliation{Department of Physics, Beijing Normal University, Beijing 100875, China}
\author{Su-Peng Kou}
\thanks{Corresponding author}
\email{spkou@bnu.edu.cn}
\affiliation{Department of Physics, Beijing Normal University, Beijing 100875, China}

\begin{abstract}
In this paper, we study the quantum properties of a bilayer graphene with
(asymmetry) line defects. The localized states are found around the line
defects. Thus, the line defects on one certain layer of the bilayer graphene
can lead to an electric transport channel. By adding a bias potential along
the direction of the line defects, we calculate the electric conductivity of
bilayer graphene with line defects using Landauer-B\"{u}ttiker theory, and
show that the channel affects the electric conductivity remarkably by
comparing the results with those in a perfect bilayer graphene. This
one-dimensional line electric channel has the potential to be applied in the
nanotechnology engineering.
\end{abstract}

\maketitle

\section{Introduction}

Recently, graphene, a two-dimensional Dirac material\cite{1,2,3,4,5,6,7}
becomes an attractive research field due to its exotic quantum properties,
such as the unconventional quantum Hall effect, the Klein paradox \cite{8},
the Josephson effect \cite{9}, and the n-p junction \cite{10}. Bilayer
graphenes are weakly-coupled two single-layer graphene by interlayer carbon
hopping, typically arranged in the Bernal (AB) stacking arrangement. In
addition to the interesting underlying physics properties, bilayer graphenes
have the potential electronics applications, owing to the possibility of
controlling both the carrier density and energy band gap through doping or
gating \cite{11,12,13,14}. A dual-gated structure allows electrical and
independent control of the perpendicular electric field and the carrier
density \cite{15,16,17}. Intrinsic bilayer graphene has no band gap between
its conduction and valence bands and the low-energy dispersion is quadratic
with massless chiral quasiparticles\cite{18,19,20}. This is in contrast to
what is observed in the monolayer which has a linear dispersion with
massless quasiparticles.
\begin{figure}[tph]
\scalebox{0.37}{\includegraphics*[0.1in,0.1in][10.7in,6.8in]{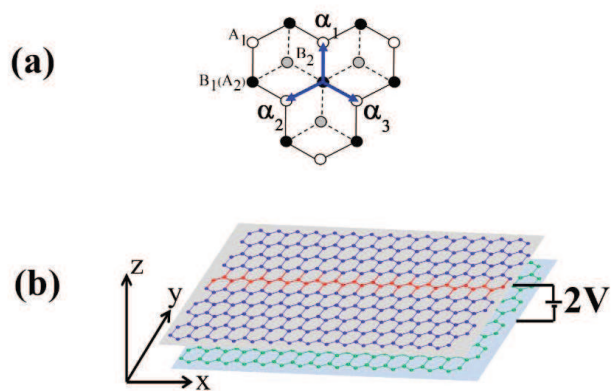}}
\caption{(Color online) \textbf{(\textbf{a})} The crystal structure of a
bilayer graphene with $A_{1},$ $A_{2},$ $B_{1},$ $B_{2}$ sublattices denoted
by white, black, black, gray circles, respectively. 1 and 2 denote upper
layer and lower layer. $\mathbf{a}_{1},\mathbf{a}_{2},\mathbf{a}_{3}$ are
primitive lattice vectors. Solid (dashed) lines denote the nearest neighbour
links on the upper (lower) layer. \textbf{(\textbf{b})} Bilayer graphene
with line defects on the upper layer marked by red lines. An electric field
is applied perpendicular to the layer (z-direction) and the corresponding
energy difference between two layers is $\left\vert 2V\right\vert $.}
\end{figure}

In this paper, we consider a bilayer graphene with asymmetry line defects
and study the defect states and the electric conductivity using the
Landauer-B\"{u}ttiker theory. This nontrivial physics properties of the
defect states may be applied to a new type of devices based on the bilayer
graphene.

The paper is organized as follows: Firstly, we introduce the tight-binding
Hamiltonian of the bilayer graphene. Secondly, we calculate the energy
structure of bilayer graphene with line lattice defects. Next, we show the
effects of the line defects on the electric conductance, including the
effects of defect-line number and the vertical electric field. Finally,
summary and conclusions are provided.

\begin{figure}[tph]
\scalebox{0.27}{\includegraphics*[0.1in,0.1in][11.7in,8.0in]{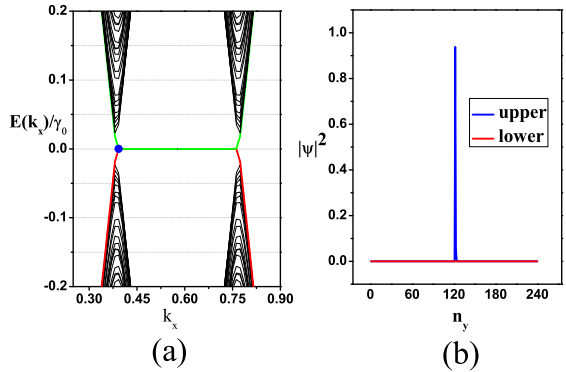}}
\caption{(Colour online) \textbf{Energy spectra and localized states with
potential}$V/\gamma _{0}=0.00$. The spectra of a single particle in the zigzag bilayer
graphene nanoribbons with a line defect located at $n_{y_{0}}=120$ on the
upper layer is shown in Fig.1(b), which has PBC along the y-direction.
\textbf{(\textbf{a})} Energy spectrum near $E(k_{x})=0$, where the red and
green lines denote the defect states of the graphene with a line-defect. The
blue dot indicates an energy state with $E(k_{x})=0$. \textbf{(\textbf{b})}
shows the particle-distribution of the zero-energy energy state indicated by
the blue dot in (a) on the upper layer (blue solid line) and lower layer(red
solid line), respectively.}
\end{figure}

\section{The tight-binding Hamiltonian}

Bilayer graphene can be classified according to the stacking type.
Generally, we focus on AB stacking, with an arrangement that was
experimentally verified in epitaxial graphene by Ohta et al \cite{21}.

Fig.1(a) is an illustration of a bilayer graphene, that consists of four
sublattices $A_{1},$ $A_{2},$ $B_{1},$ $B_{2}$ and the lattice constant is
set to unit. To study the system simply, we mainly focus on the
non-interaction case, of which an AB-stacked bilayer graphene model only
contains first nearest-neighbor (NN) intralayer hopping and NN interlayer
hopping. The tight-binding Hamiltonian is given by
\begin{equation}
H_{0}=-\gamma _{0}\sum_{l,\langle i,j\rangle }\left( a_{l,i}^{\dag
}b_{l,j}+h.c.\right) -\gamma _{1}\sum_{i}\left( a_{1,i}^{\dag
}b_{2,i}+h.c.\right) .
\end{equation}
Here, $a_{l,i}~(b_{l,i})$ is the annihilation operator at sublattice $%
A_{l}~(B_{l})$ at site $\mathbf{R}_{i},$ where $l=1$ represents the upper
layer and $l=2$ represents the lower layer of the bilayer material. The
first term in the Hamiltonian describes intralayer nearest-neighbor coupling
inside the layer, the second term describes the interlayer coupling between the sublattice of the lower layer $B_{2}$ and the sublattice of the upper layer $A_{1}$ and here we set $\gamma _{0}=3.16$ eV, $\gamma _{1}=0.39$ eV) \cite{19.1}.
\begin{figure}[tph]
\scalebox{0.27}{\includegraphics*[0.1in,0.1in][11.7in,8.0in]{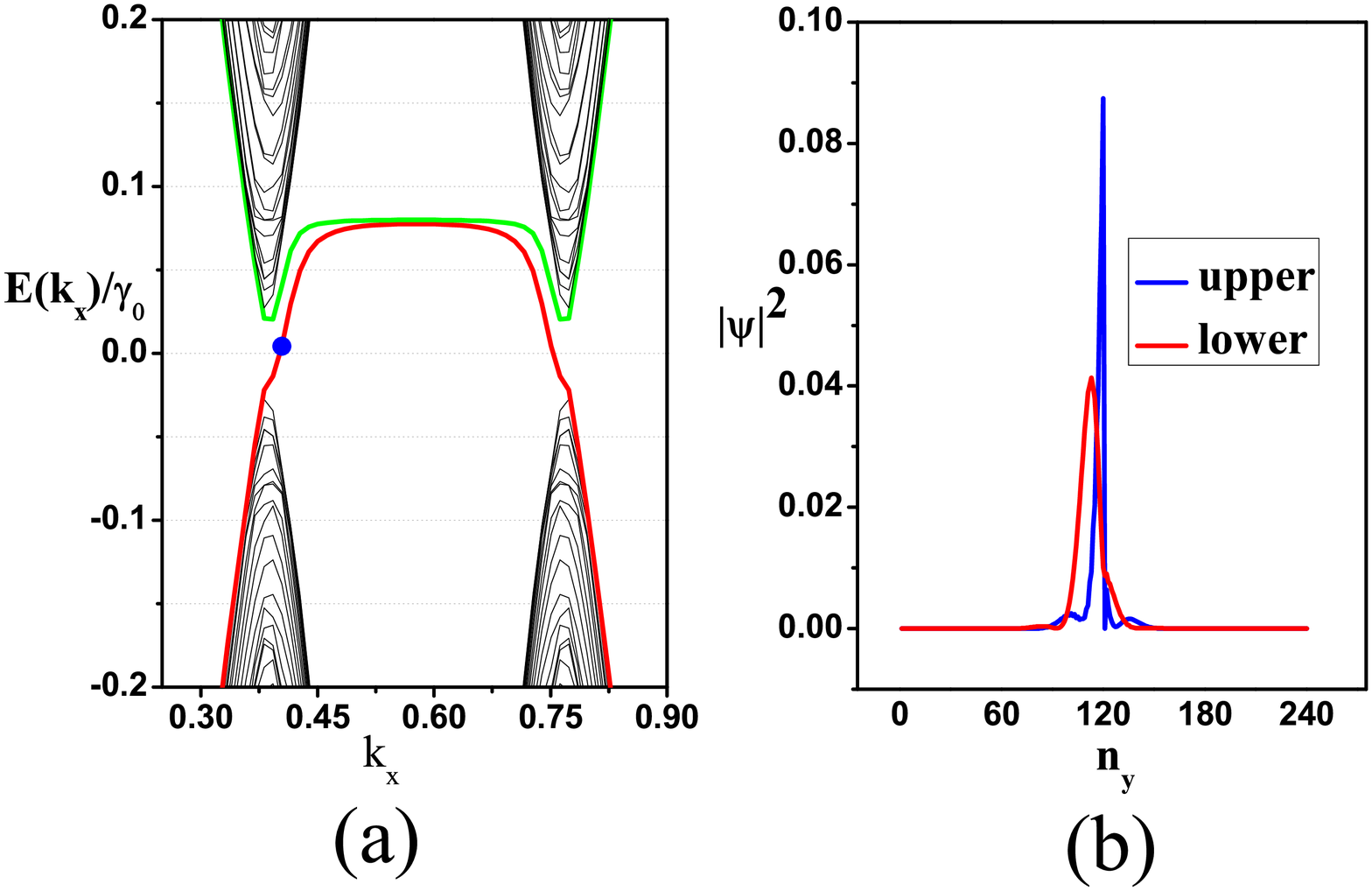}}
\caption{(Colour online) \textbf{Spectra of localized states with potential
}$V/\gamma _{0}=0.08$. Except for the potential, other conditions are the same as in
Fig.2. The wave function in (b) corresponds to the energy state indicated by
the blue dot in (a). In this case, an energy gap opens and the dispersion of
the defect state has no flat band but looks like an inverted "$U$". }
\end{figure}

Due to spacial translation symmetry, intrinsic bilayer graphene can be
described in momentum space and the number of sites in a primitive cell is
four. Owing to the inversion symmetry in neutral bilayer graphene, there
exists the degeneracy of the highest valence and lowest conduction bands. If
the inversion symmetry is broken, an mass gap opens in the low energy
spectrum \cite{22,23,24}. When we assume that the upper and lower layers are
at different electrostatic potentials $V$ (normally called bias potentials),
the inversion symmetry is broken. Hence, the energy difference between the
two layers is parameterized by the energy $V$. Then, the Hamiltonian becomes
\begin{align}
H=& H_{0}+H_{V}  \notag \\
=& -\gamma _{0}\sum_{l,\langle i,j\rangle }\left( a_{l,i}^{\dag
}b_{l,j}+h.c.\right) -\gamma _{1}\sum_{i}\left( a_{1,i}^{\dag
}b_{2,i}+h.c.\right)   \notag \\
+& \sum_{l,\langle i,j\rangle }(V_{l,i}a_{l,i}^{\dag
}a_{l,i}+V_{l,j}b_{l,j}^{\dag }b_{l,j}),
\end{align}%
where $H_{V}$ is the external electrostatic potential. Here, we consider a
staggered energy potential for bilayer graphene with $V_{1,i}=V$ on the
upper layer and $V_{2,i}=-V$ on the lower layer, where $V\geq 0$ is a
controllable constant quantity.

\section{Localized states induced by line defects}

In this paper, we assume that a line-lattice defect lies on the upper layer,
as represented by the highlighted red line marked in Fig.1(b), which is a
zigzag lattice chain. To study the line defects on bilayer graphene, we
simulate lattice defects on bilayer graphene by the following method: for a
target lattice site $i$ on line defects, we set all the coupling hopping
parameters connected to the site $i$ to $0$ and the local on-site energy
potential $V_{i}$ to be a relatively larger value. To study the energy
spectrum we assume the bilayer graphene is semi-infinite, that is, the
system is finite along the $y$-direction and infinite along the $x$%
-direction. Thus, the line defect is an infinite zigzag chain along the $x$%
-direction. To study the localized states induced by line defects, we use
\begin{figure}[tph]
\scalebox{0.27}{\includegraphics*[0.1in,0.1in][11.7in,8.0in]{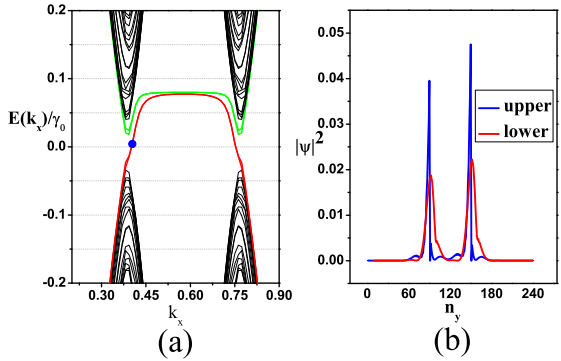}}
\caption{(Colour online) \textbf{Energy spectra and localized states when
}$V/\gamma _{0}=0.08$. The wave function in (b) corresponds to the quantum state of the
energy indicated by the blue dot in (a). Here are two parallel line defects
at $n_{y}=80$ and $n_{y}=160$ on the upper layer. }
\end{figure}

the period boundary conditions (PBC) along the $y$-direction which makes the
system a nanotube. By performing Fourier analysis and diagonalization, we
may obtain the eigenvalues (or energy spectrum) $E(k_{x})$ and the
corresponding eigenstates. The effects of the line defects and vertical
electric potential are shown in Fig.2, Fig.3, Fig.4, respectively. For the
case of $V>0$, there exists bulk energy gap, and the value of the gap
becomes larger with an increase of $V$. In particular, a line defect leads
to localized states (or defect states).

We consider a line defect on the upper layer and set the total number of
zigzag chain (parallel to the red zigzag line in Fig.1(b)) along $y$%
-direction to be $N_{y}=240$. When $V=0.00$ there exist defect states
(marked by the red solid line and green solid line in Fig.2(a)) and
zero-energy states, which has a flat band. For this case, the wave function
of zero-energy state is localized around the defects (Fig.2(b)). The
length-scale of the localized states along $x$-direction is $\xi \sim
1/\Delta _{f}$ where $\Delta _{f}$ is the energy gap of bulk system. For the
case of $V>0$, we also have the localized states around line defects. For
this case, a bulk gap appears and the defect states becomes dispersive. The
probability density of zero-energy wave function $\left\vert \psi
_{E=0}(y)\right\vert ^{2}$ becomes extended on both upper layer and lower
layer of the bilayer graphene, and the maximum value of $\left\vert \psi
_{E=0}(y)\right\vert ^{2}$ is on the corresponding positions on lower layer.
\begin{figure}[tph]
\scalebox{0.27}{\includegraphics*[0.1in,0.1in][10.5in,4.0in]{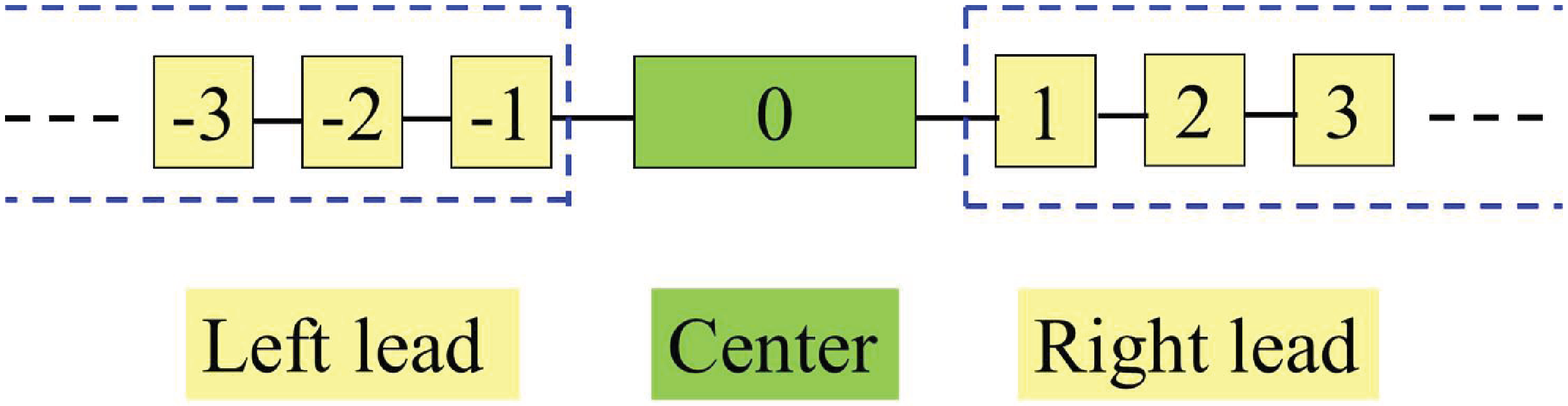}}
\caption{(Color online) The illustration of the system to calculate the
electric conductance, which consist of center device area (denoted by $0$
with green area) and left/right leads (denoted by other integer numbers with
yellow area). The structure of center can be various, but here we focus on
the bilayer graphene with line defects shown in Fig.1(b).}
\end{figure}

Furthermore, we can also study the cases with several parallel line defects.
For two parallel line defects on the upper layer, there exists localized
defect states which are located around corresponding line defects, and the
electric field induces a finite bulk energy gap (Fig.4).

\section{Transport properties of a bilayer graphene with asymmetry line
defects}

Bilayer graphene exhibits nontrivial transport properties owing to its
unusual band structure where the conduction and valence bands touch with
quadratic dispersion. One of the earliest theoretical papers studying the
conductivity through AB-stacked bilayer graphene was Ref. \cite{25} where it
was assumed that the band structure of the bilayer is described by two bands
closest to the Dirac point energy. Transport properties and the nature of
conductivity near the Dirac point were probed experimentally \cite{1,26} and
investigated theoretically \cite{27}.

We use the well-known Landauer-B\"{u}ttiker equation to obtain the electric
conductivity. To show the effect of the line defect on the electric
conductance, we seperate the original system into three parts: left lead,
right lead and the center device area, as shown in the schematic diagram in
Fig.5. The left and right leads are semi-infinite and the center device area
has finite size. The number of primitive cells along the $x$-direction is $%
N_{x}$; the number of zigzag chain along the $y$-direction is $N_{y}$. Here,
it is a zigzag edge along the $x$-direction and an armchair edge along the $%
y $-direction. Hence, for the lead-center-lead bilayer graphene system, the
zero-temperature conductance is calculated using the Landauer-B\"{u}ttiker
formalism \cite{28}:%
\begin{equation}
G=\frac{2e^{2}}{h}\sum_{\alpha ,\beta }(T)_{\beta \alpha },
\end{equation}%
\begin{figure}[tph]
\scalebox{0.30}{\includegraphics*[0.1in,0.1in][11.7in,8.0in]{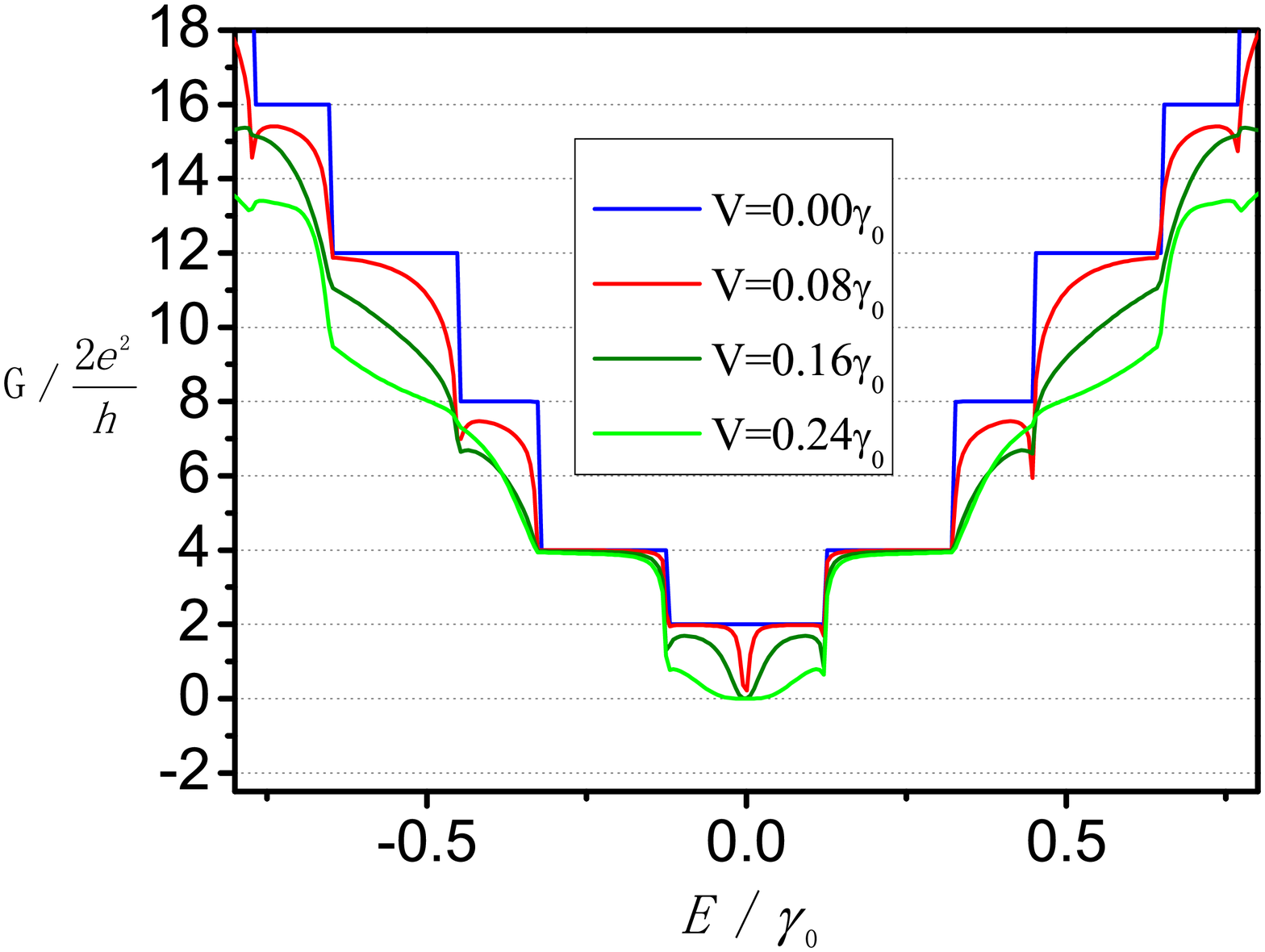}}
\caption{(color line) \textbf{The electric conductance with different
electric potential $V$.} This is based on a perfect bilayer graphene device
with PBC, i.e., bilayer graphene nanotube. When $V/\gamma _{0}=0.00$, the minimum value
of G is $2(\frac{2e^{2}}{h})$}
\end{figure}
where $(T)_{\beta \alpha }$ is the transmission from the incoming state $%
\alpha $ in the left lead to the outgoing state $\beta $ in the right lead.
The terminal electric conductance can be written as $G=\frac{2e^{2}}{h}T,$
and the transmission probability $T$ through it is
\begin{equation}
T=Tr[\mathbf{\Gamma }_{L}\mathbf{G}_{C}^{r}\mathbf{\Gamma }_{R}\mathbf{G}%
_{C}^{a}],
\end{equation}%
where $\mathbf{\Gamma }_{L}$ $(\mathbf{\Gamma }_{R})$ is the left (right)
coupling matrix, and $\mathbf{G}_{C}^{r}$ $(\mathbf{G}_{C}^{a})$ is the
retarded (advanced) Green's function of the center part of the system.

The single particle Green's function of the center devise area $G_{c}$ is
defined as
\begin{equation}
G_{c}=[(E+i\eta )\mathbf{I}-\mathbf{H}_{c}-\Sigma _{L}-\Sigma _{R}]^{-1}
\end{equation}%
where $E$ is the energy of the particle, $\eta $ is a positive
infinitely-small real number, $\mathbf{I}$ represents the identity matrix,
and $\mathbf{H}_{c}$ is the Hamiltonian of the center devise area. $\Sigma
_{L}$ ($\Sigma _{R}$) are the contact self-energies reflecting the effect of
coupling of the center part of the graphene to the left (right) leads. We
obtain the expressions for the self-energies of the two leads as
\begin{align}
\mathbf{\Sigma }_{R}& \mathbf{=}\mathbf{H}_{0,1}\mathbf{g}_{1,1}^{R}\mathbf{H%
}_{1,0},  \notag \\
\mathbf{\Sigma }_{L}& \mathbf{=}\mathbf{H}_{0,-1}\mathbf{g}_{-1,-1}^{L}%
\mathbf{H}_{-1,0},
\end{align}%
where $\mathbf{H}_{0,1}$ $\mathbf{(H}_{0,-1}\mathbf{)}$ is the coupling
matrix of the center part to the right (left) lead. Subsequently, we can
successfully calculate the electron transmission probability using $\mathbf{H%
}_{1,1},$ $\mathbf{H}_{1,2}\mathbf{,}$ $\mathbf{H}_{1,0}$. $\mathbf{g}%
_{1,1}^{R}(\mathbf{g}_{-1,-1}^{L})$ are the surface Green's functions of the
right (left) leads. Hence, we can determine the conductivity in the presence
and absence of line defects.
\begin{figure}[tph]
\scalebox{0.30}{\includegraphics*[0.1in,0.1in][11.7in,8.0in]{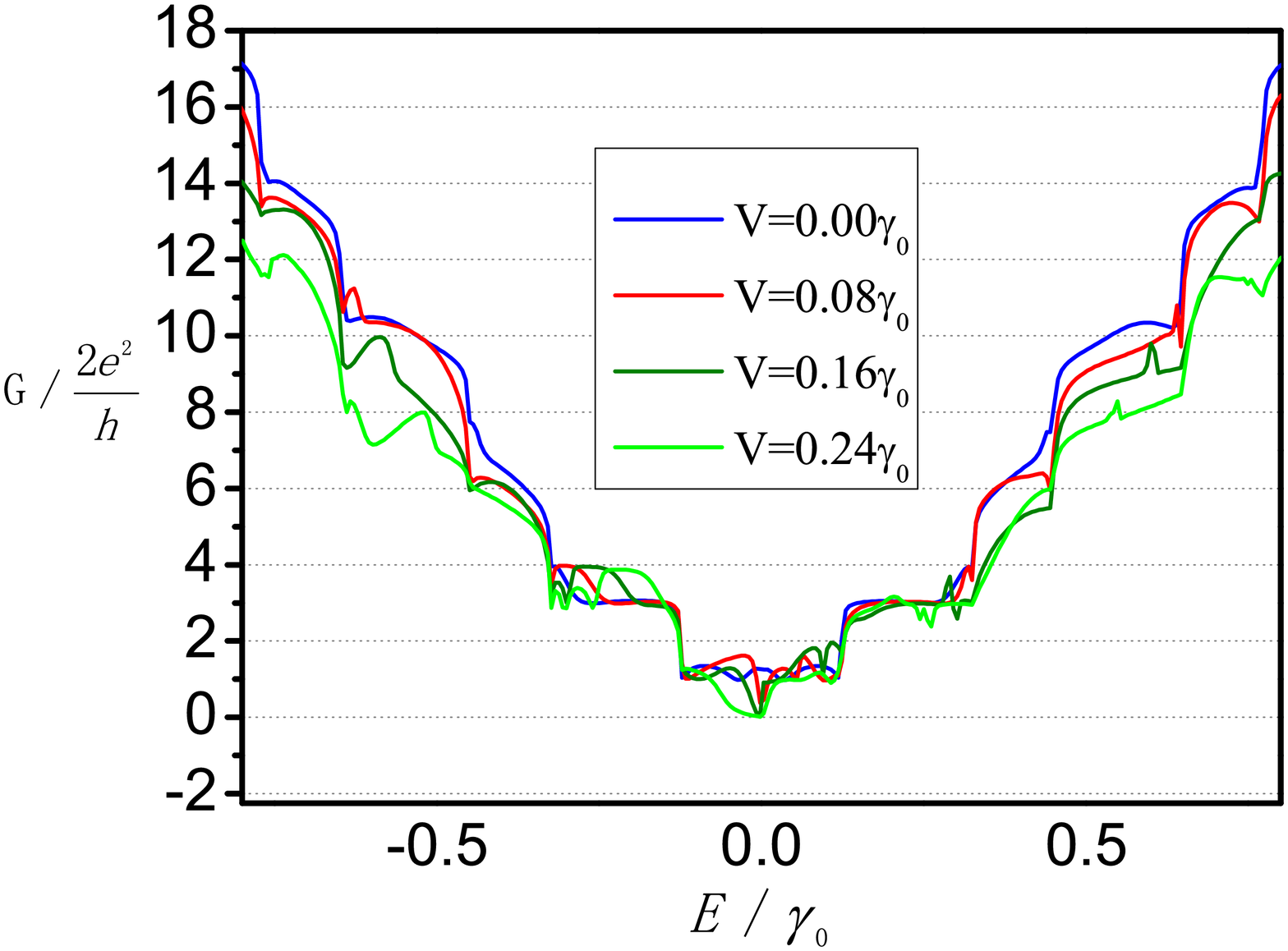}}
\caption{(color online) \textbf{The electric conductance with different
electric potential $V$.} This is based on a line-defected bilayer graphene
device with PBC along $y$-direction.}
\end{figure}

To compare with the electrical transport effects of the line defects, the
electric conductance of perfect bilayer graphene devices is also given
(Fig.6). We use PBC along the $y$-direction, that is, the center devices
area is a bilayer graphene nanotube. When $V=0$, the value of $G$ near $E=0$
is $2(\frac{2e^{2}}{h})$, which is double to the value of one layer graphene
$\frac{2e^{2}}{h}$ owing to the degeneracy of the highest valence and lowest
conduction bands. When $V>0$, the value of $G$ at $E=0$ approaches zero,
which indicates that the system generates a bulk energy gap, and the larger $%
V$ results in a larger width of gap. However, the electric conductance shows
additional features induced by the line defects as shown in Fig.7. When $V=0$
the value of $G$ near $E=0$ is about $\frac{2e^{2}}{h}$. When $V>0$
there is $G(E)\neq G(-E)$, and from the contrast of pure bilayer graphene and symmetric defects it is clear that this is the result of the particle-hole symmetry breaking by the asymmetric defects \cite{29,30}. Now, the degeneracy of the zero energy is left and the energy
spectrum becomes asymmetry. Hence, the conductance can be controlled by the
vertical electric potential and the line defects in this case. As a result,
this one-dimensional line electric channel has the potential to be applied
in the nanotechnology engineering.

\section{Conclusion}

In this paper, we studied the physics properties of a bilayer graphene with
line defects, including the defect-induced localized states and the electric
conductivity. We found that the line defect on a certain layer of the
bilayer graphene leads to an electric channel. When $V=0$, the localized
states on single layer has a flat band with zero energy. When $V>0$ the
system becomes gapped and the localized modes have the distribution on both
layers. The additional conductance $G$ from line defect is obtained. This
effect from line defects may be applid in electronic devices based on
bilayer graphene.

\begin{acknowledgments}
This work is supported by National Basic Research Program of China
(973Program) under the grant No. 2011CB921803, 2012CB921704 and NSFC
GrantNo.11174035, 11474025, 11504285, 11404090 and SRFDP, the Fundamental
Research Funds for the Central Universities, and the Scientific Research
Program Funded by Shanxi Provincial Education Department under the grant No.
15JK1363.
\end{acknowledgments}

\end{document}